\documentclass[journal]{IEEEtran}
\usepackage{cite}

\ifCLASSINFOpdf
  \usepackage[pdftex]{graphicx}
\else
\fi
\usepackage{amsmath}
\usepackage{amsfonts}
\usepackage{algorithmic}

\usepackage{array}

\usepackage[caption=false,font=normalsize,labelfont=sf,textfont=sf]{subfig}

\usepackage{stfloats}
\usepackage{url}

\begin{document}
%
\title{Nonlinear Stability Boundary Assessment of Multi-Converter Systems Based On Reverse Time Trajectory}
\author{Sujay~Ghosh, Mohammad Kazem Bakhshizadeh,  Guangya Yang and~Łukasz Kocewiak
\thanks{S. Ghosh (corresponding author), M. K. Bakhshizadeh, and Ł. Kocewiak are with Ørsted Wind Power, Nesa Allé 1, 2820, Denmark. (e-mail: sujgh@orsted.com; modow@orsted.com;  lukko@orsted.com)}
\thanks{G. Yang is with the Technical University of Denmark, Anker Engelunds Vej 1, 2800, Denmark (e-mail: gyy@elektro.dtu.dk).}
\thanks{Manuscript received xx xx xxxx; revised xx xx xxxx.}}

\markboth{Journal of \LaTeX\ Class Files,~Vol.~14, No.~8, August~2015}%
{Shell \MakeLowercase{\textit{et al.}}: Bare Demo of IEEEtran.cls for IEEE Journals}

\maketitle

\begin{abstract}
As the integration of wind power accelerates, wind power plants (WPPs) are expected to play a crucial role in ensuring stability in future power grids. This paper examines the nonlinear stability boundary of a multi-converter system in a wind power plant (WPP) connected to an AC power grid via a long HVAC cable. Traditionally, for nonlinear analysis of WPPs, a simplification is adopted wherein the WPP is treated as an aggregation of individual wind turbines (WTs), with a simplified portrayal of the collector network. However, in the presence of different technologies, such as STATCOM, that are placed away from the WTs, the model aggregation will not hold. This paper presents a unified methodology to model and investigate the high-dimensional stability boundary of a WPP with a STATCOM. The stability region of the system, i.e. the region of attraction (RoA), is determined by the reverse time (backwards) trajectory technique. Furthermore, the estimated stability boundary is verified using time-domain simulation studies in PSCAD.
\end{abstract}


\IEEEpeerreviewmaketitle

\section{Introduction}
\IEEEPARstart{I}{n} recent times, the energy sector has undergone a significant transformation, primarily influenced by advancements in renewable energy technologies, particularly wind power and photovoltaic (PV) power generation. This evolution has led to a gradual increase in the percentage of renewable energy sources within the power grid. Notable forecasts indicate that worldwide wind power integration envisaged by the Net Zero Scenario calls for an average expansion of approximately 17\% per year during 2023-2030 \cite{1}. As the integration of wind power accelerates, wind power plants (WPPs) are expected to play a crucial role in ensuring stability in future power grids. 

Navigating power system stability and control challenges requires accurate system modelling and relevant stability assessment methodologies. Linearised model-based approaches, such as eigenvalue analysis or impedance-based stability analysis, assume the system has a linear behaviour under small disturbances, and synchronisation is enforced only within the vicinity of the operating point \cite{2}\cite{3}. Conversely, the nonlinear approach to stability includes large-signal disturbance and can provide global asymptotic stability conclusions \cite{4}\cite{5}.

A typical wind power plant consists of several wind turbines (WTs). Traditionally, for analysis of WPPs, a simplification is adopted wherein a WPP is treated as an aggregation of individual wind turbines \cite{6}\cite{7}. However, only in \cite{8} was it presented that WTs can be aggregated only up to a certain frequency range, i.e. up to 400 Hz. In this regard, in \cite{9}, a reduced-order representation of a WPP was discussed, and it was demonstrated that modelling only the low-frequency dynamics, such as synchronisation controls, fault-ride through controls, etc., is sufficient to investigate the large disturbance/ nonlinear stability of the system. However, the existing literature still does not comprehensively discuss a unified methodology to investigate the nonlinear stability with different technologies, such as WTs with STATCOM. 

The WTs are geographically dispersed and interconnected through a collector network to the AC grid. Various studies have probed aspects like small-signal stability and interactions among WTs within the WPP context \cite{10}\cite{11}. However, certain studies' oversimplified portrayal of collection lines \cite{11} might lack fidelity in capturing real-world complexities. Recent literature like \cite{12} has proposed a WPP collector network modelling approach that acknowledges the intricacies of collector lines. Nevertheless, these models prove inadequate for investigating the dynamic behaviours of the WPP. 

Consequently, this paper aims to bridge a critical research gap by introducing a comprehensive WPP model incorporating the WTs, the collector network and possible shunt compensation devices like a STATCOM suitable for investigating the nonlinear stability of WPP connections. This work employs the reverse-time trajectory technique, introduced in \cite{13}, to estimate a time-limited region of attraction of the system. This innovative technique facilitates quick estimation of the system RoA, without being too optimistic or too conservative. In summation, this study responds to the pressing need for a comprehensive methodology to study the nonlinear stability of a multi-converter system in a WPP, paving the way for more effective and informed decision-making in the pursuit of renewable energy integration.

The main contribution of this paper is as follows, 
\begin{itemize}
    \item A systematic approach to modelling a full WPP incorporating the WTs, the collector network and possible shunt compensation devices like a STATCOM suitable for investigating the nonlinear stability of WPP connections is discussed.

    \item The concept of a multi-dimensional region of attraction for the multi-converter system in a WPP is introduced. A technique to visualise the multi-dimensional region of attraction in a lower dimension is also discussed. 
\end{itemize}

The rest of the paper is organised as follows: Section II presents the comprehensive modelling of the WPP. Section III presents the methodology to access the nonlinear stability of the WPP connection. In Section IV, the estimated nonlinear boundary is validated against a detailed WT simulation model in PSCAD. Finally, Section V presents the conclusion.

\section{System Modelling}
The WPP considered in this work is shown in Fig. 1. The WPP can be divided into three parts: the WTs, the STATCOM, and the collector network connected to the AC grid. Without loss of generality, the WTs are type-4. As discussed in Section I, an aggregated reduced order model (ROM) of the WT is considered since, for nonlinear stability, only the low-frequency dynamics are sufficient. The collector network in the WPP is radial, where all the collector lines converge to the POC point (i.e. 66kV) and finally connect to the grid through the substation.

\begin{figure}[h]
    \centering
    \includegraphics[width=9.0cm]{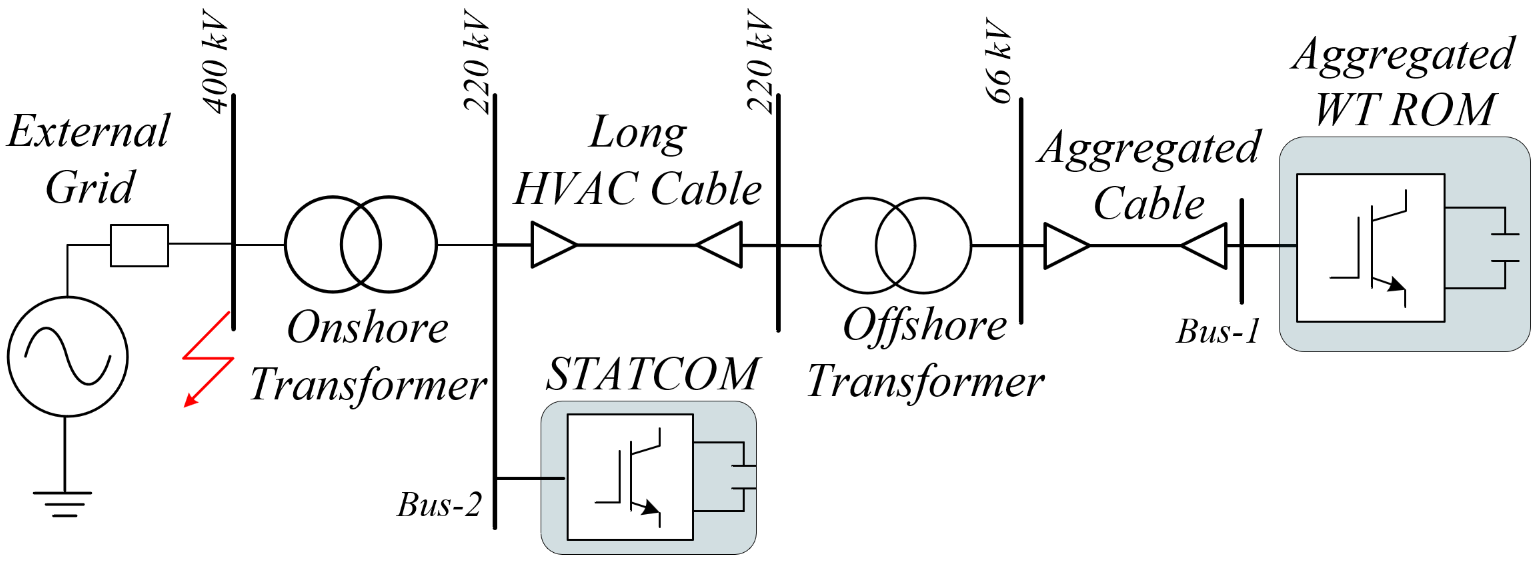}
    \caption{WPP with aggregated WTs and a STATCOM, connected to an AC power grid.}
    \label{ROM}
\end{figure}

\begin{figure*}[h]
    \centering
    \includegraphics[width=15.20cm]{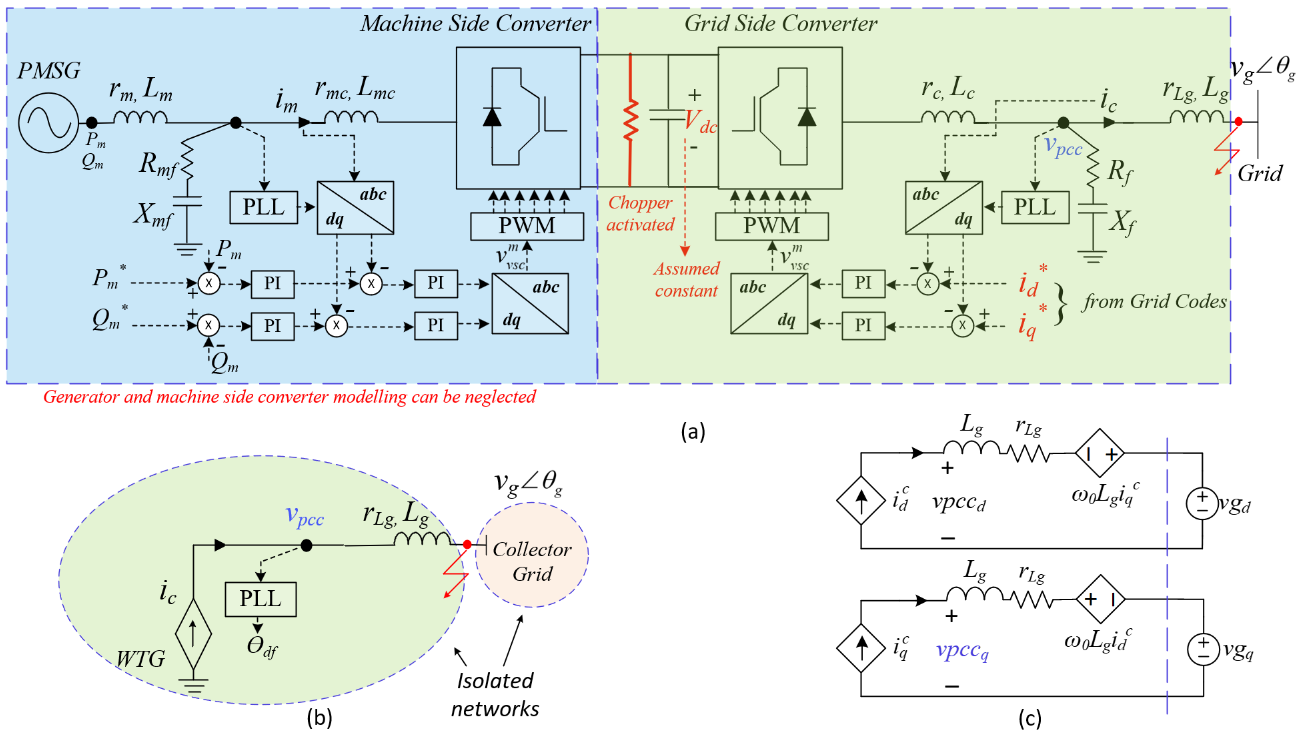}
    \caption{Wind turbine model: (a) Reduced order model (ROM) of the Type-4 wind turbine considering the actions/assumptions. (b) System representation of ROM in the DQ domain.}
    \label{ROM}
\end{figure*}

\subsection{Modelling of reduced-order type-4 WTs}
The structure and the control system of a type-4 WT are shown in Fig. 2. This paper adopts a reduced second-order model of the type-4 WT, considering that the current control dynamics are much faster than that of the PLL. 

The equivalent mathematical model of the WT unit \cite{14}\cite{15} can be presented as,
\begin{equation}\label{SEq_1}
\begin{aligned}
\dot{x_1} &= x_2 \\
x_{2} &= x_{2}^{max} \cdot \text{tanh}( {x_{3}}/{x_{2}^{max}})\\
M_{wt} \dot{x_3} &= T_{m_{wt}} - T_{e_{wt}} - D_{wt} {x_2}\\
\end{aligned}
\end{equation}

where, $x_1= \theta_{wt,pcc}$, $x_2= \dot{\theta}_{wt,pcc}$ (saturated), and $x_3= \dot{\theta}_{wt,pcc}$ (unsaturated),
\begin{equation}\label{SEq_2}
\begin{aligned}
M_{wt} &= 1- k_p L_g i_d^c\\
T_{m_{wt}} 
&= k_p( \dot{\overline{r_{Lg} i_q^c}} + \ddot{\overline{L_g i_q^c}} + \dot{\overline{L_g i_d^c}} \omega_g)
+ k_i( r_{L_g} i_q^c \\
&\qquad+ \dot{\overline{L_g i_q^c}} + L_g i_d^c \omega_g) \\
T_{e_{wt}} 
&=  k_i V_g \text{sin} (x_1 - \theta_{g}) + k_p \dot{V_g} \text{sin} (x_1 - \theta_{g}) + M_{wt} \dot{\omega}_g\\
D_{wt} 
&= k_p ( V_g \text{cos}(x_1 - \theta_{g}) - \dot{\overline{L_g i_d^c}}) - k_i L_g i_d^c
\end{aligned}
\end{equation}

The $r_{Lg}$ and $L_g$ are the WT transformer resistance and inductance, respectively; $i_d^c$ and $i_q^c$ are the WT currents in the converter reference frame; and $V_g$ and $\theta_g$ is the grid voltage in the system reference frame. $k_p$ and $k_i$ are gains of the PLL.

The system (1) is modelled in a DQ frame rotating at a fixed frequency $\omega_0$.  During grid faults, the current injection $i_d^c$ and $i_q^c$ can be computed as (3), where, $i_{q,3ph}^c$ is limited to 1 pu.
\begin{equation}\label{}
\begin{aligned}
i_{q,3ph}^c &= K_{factor} \cdot v_{pcc} \\
i_{d,3ph}^c &= \sqrt{I_{max}^2 - (i_{q,3ph}^c)^2} \\
\end{aligned}
\end{equation}
where, $v_{pcc}$ is the voltage at point of common coupling; and $I_{max}$ is the maximum current rating of the converter.
Further, based on aggregation characteristics, considering $N$ turbines in the WPP, the system (1) can be scaled up as, 
\begin{equation}\label{}
\begin{aligned}
i_d^c &= (N \cdot i_d^c), i_q^c = (N \cdot i_q^c)\\
r_{Ls} &= (r_{Ls} / N), L_{s} = (L_{s} / N). 
\end{aligned}
\end{equation}

Note, WTs can be aggregated only up to a frequency range of 400 Hz \cite{8}, therefore our analysis is limited to the same.

\subsection{Modelling of STATCOM}
A grid-following STATCOM can be modelled similarly to a WT ROM, as described in Section II-A. However, for a STATCOM, the active current injection is zero. Under normal operation, the reactive current $i_{st,q}^c$ is injected based on system requirements. However, during severe grid faults, $i_{st,q}^c=1pu$. Therefore, the equivalent mathematical model for the STATCOM can be presented as,

\begin{equation}\label{SEq_1}
\begin{aligned}
\dot{y_1} &= y_2 \\
y_{2} &= y_{2}^{max} \cdot \text{tanh}( {y_{3}}/{y_{2}^{max}})\\
M_{st} \dot{y_3} &= T_{m_{st}} - T_{e_{st}} - D_{st} {y_2}\\
\end{aligned}
\end{equation}

where, $y_1= \theta_{st,pcc}$, $y_2= \dot{\theta}_{st,pcc}$ (saturated), and $y_3= \dot{\theta}_{st,pcc}$ (unsaturated)
\begin{equation}\label{SEq_2}
\begin{aligned}
M_{st} &= 1\\
T_{m_{st}} 
&= k_p( \dot{\overline{r_{Ls} i_{st,q}^c}} + \ddot{\overline{L_s i_{st,q}^c}})
+ k_i( r_{L_s} i_{st,q}^c + \dot{\overline{L_s i_{st,q}^c}}) \\
T_{e_{st}} 
&=  k_i V_s \text{sin} (y_1 - \theta_{s}) + k_p \dot{V_s} \text{sin} (y_1 - \theta_{s}) + \dot{\omega}_s\\
D_{st} 
&= k_p  V_s \text{cos}(y_1 - \theta_{s})
\end{aligned}
\end{equation}

The $r_{Ls}$ and $L_s$ are the STATCOM transformer resistance and inductance, respectively; $i_{st,q}^c$ is the STATCOM reactive current in the converter reference frame; and $V_s$ and $\theta_s$ is the grid voltage in the system reference frame. $k_p$ and $k_i$ are gains of the STATCOM PLL.

\subsection{Modelling of collector network}
As the scale of WPP increases, the topology and impedance of the collector network become non-negligible, especially when studying the internal dynamics of the WPP. The structure of the WPP can be re-visualised as in Fig. 3. In this work, the impedance of the WPP collector network is proposed to be obtained by a multi-port frequency scan, looking from the aggregated WT bus and the STATCOM bus. 

\begin{figure}[h]
    \centering
    \includegraphics[width=4.50cm]{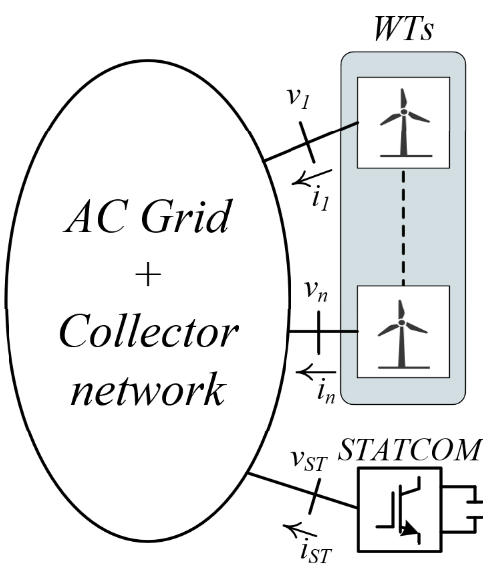}
    \caption{Re-visualisation of wind power plant for multi-port frequency scan.}
    \label{ROM}
\end{figure}

A multi-port frequency scan, looking from WT bus-1 and STATCOM bus-2, can be presented as,

\begin{equation}\label{}
\begin{aligned}
Z_{WPP} = 
\begin{bmatrix}
Z_{11}(s) & Z_{12}(s)\\
Z_{21}(s) & Z_{22}(s)
\end{bmatrix}
\end{aligned}
\end{equation}

Furthermore, as suggested in \cite{9}, for nonlinear stability analysis, the impedance scan of the collector network can be fitted around $\pm$5 Hz of the PLL frequency due to the hard saturation limits. A simplified curve-fitting equation is as follows,  
\begin{equation}\label{}
\begin{aligned}
r_{ij} &= \mathbb{R}\{Z_{ij}(F_{C})\} \\
L_{ij} &= m/ (2\pi) \\
\end{aligned}
\end{equation}
where, $Z_{ij}(f)$ is the impedance-frequency scan vector, $F_C$ is the corner frequency of the frequency scan, and $m$ is the slope of the frequency scan around the $\pm$5 Hz of the PLL frequency. For more information, refer to \cite{9}.  
Once the $r_{ij}$ and $L_{ij}$ are estimated, the equation governing the collector network can be written as (9).

\begin{equation}\label{vpcc-1}
\begin{aligned}
\begin{bmatrix}
v_{d, i}^s\\
v_{q, i}^s
\end{bmatrix} &=
 \left[\begin{matrix}
  v0_{d,i}\\
  v0_{q,i}
\end{matrix}\right] + \left[\begin{matrix}
  L_{ii} s + r_{ii}  & -\omega_{0} L_{ii}\\
  \omega_{0} L_{ii} & L_{ii} s + r_{ii}
\end{matrix}\right] \left[\begin{matrix}
  i_{d,i}^s\\
  i_{q,i}^s
\end{matrix}\right]\\
&\qquad + \left[\begin{matrix}
  L_{ij} s + r_{ij}  & -\omega_{0} L_{ij}\\
  \omega_{0} L_{ij} & L_{ij} s + r_{ij}
\end{matrix}\right] \left[\begin{matrix}
  i_{d,j}^s\\
  i_{q,j}^s
\end{matrix}\right]\\
\end{aligned}
\end{equation}
The inputs to (9) are the open circuit WT and STATCOM bus voltages, and current injection in the system reference frame. The outputs are the bus voltages in the system reference frame. 
The RHS of (9) can be simplified as (10) by simply solving the derivatives.

\begin{equation}\label{vpcc-1}
\begin{aligned}
\begin{bmatrix}
v_{d, 1}^s\\
v_{q, 1}^s
\end{bmatrix} &=
 \left[\begin{matrix}
  V_{1,0} $cos$ (\theta_{g,0})\\
  V_{1,0} $sin$ (\theta_{g,0})
\end{matrix}\right]\\
&+\left[\begin{matrix}
  r_{11}  & -\omega_{0} L_{11}\\
  \omega_{0} L_{11} & r_{11}
\end{matrix}\right] \left[\begin{matrix}
  $cos$ (x_1) & -$sin$ (x_1)\\
  $sin$ (x_1) & $cos$ (x_1)
\end{matrix}\right] \left[\begin{matrix}
  i_{d}^{c}\\
  i_{q}^{c}
\end{matrix}\right]\\
&+
\frac{d(x_1)}{dt} \left[\begin{matrix}
  L_{11} & 0\\
  0 & L_{11}
\end{matrix}\right] \left[\begin{matrix}
  -$sin$ (x_1) & -$cos$ (x_1)\\
   $cos$ (x_1) & -$sin$ (x_1)
\end{matrix}\right] \left[\begin{matrix}
  i_{d}^{c}\\
  i_{q}^{c}
\end{matrix}\right]\\
&+
\left[\begin{matrix}
  L_{11} & 0\\
  0 & L_{11}
\end{matrix}\right] \left[\begin{matrix}
  $cos$ (x_1) & -$sin$ (x_1)\\
  $sin$ (x_1) &  $cos$ (x_1)
\end{matrix}\right] \frac{d}{dt}\left[\begin{matrix}
  i_{d}^{c}\\
  i_{q}^{c}
\end{matrix}\right]\\
&+\left[\begin{matrix}
  r_{12}  & -\omega_{0} L_{12}\\
  \omega_{0} L_{12} & r_{12}
\end{matrix}\right] \left[\begin{matrix}
  $cos$ (y_1) & -$sin$ (y_1)\\
  $sin$ (y_1) &  $cos$ (y_1)
\end{matrix}\right] \left[\begin{matrix}
  0\\
  i_{st,q}^{c}
\end{matrix}\right]\\
&+
\frac{d(y_1)}{dt} \left[\begin{matrix}
  L_{12} & 0\\
  0 & L_{12}
\end{matrix}\right] \left[\begin{matrix}
  -$sin$ (y_1) & -$cos$ (y_1)\\
  $cos$ (y_1) & -$sin$ (y_1)
\end{matrix}\right] \left[\begin{matrix}
  0\\
  i_{st,q}^{c}
\end{matrix}\right]\\
&+
\left[\begin{matrix}
  L_{12} & 0\\
  0 & L_{12}
\end{matrix}\right] \left[\begin{matrix}
  $cos$ (y_1) & -$sin$ (y_1)\\
  $sin$ (y_1) &  $cos$ (y_1)
\end{matrix}\right] \frac{d}{dt}\left[\begin{matrix}
  0\\
  i_{st,q}^{c}
\end{matrix}\right]\\
\end{aligned}
\end{equation}

Equation (10) is derived for calculating WT bus voltages. The STATCOM bus voltage [$v_{d, 2}^s$; $v_{q, 2}^s$] can be derived similarly; in order to avoid repetition, it is not presented. It must be noted that the reduced order model (1)-(10) has been derived based on differential equations without any RMS assumptions.

\section{Nonlinear Stability Assessment}
This section presents the nonlinear stability assessment methodology of the multi-converter system described in Fig. 1. Table 1 and Table 2 present the operating point of the WT converter system and the STATCOM, respectively. The modelling parameters of the collector network, transformers and export cable are described in detail in \cite{16}. 

\begin{table}[h]
\caption{WT SYSTEM AND CONTROL PARAMETERS}
\label{table}
\setlength{\tabcolsep}{3pt}
\begin{tabular}{|p{40pt}|p{105pt}|p{70pt}|}
\hline
Symbol& 
Description& 
Value \\
\hline
$S_b $& 
Rated power& 
20x12 MVA \\
$V_g$& 
Nominal grid voltage (L-N, pk) & 
690 $\sqrt{2/3}$ V\\
$f_g$& 
Rated frequency & 
50 Hz \\
$r_{Lg}$, $L_{g}$& 
Transformer impedance & 0.25m$\Omega$, 0.018mH
\\
$i_d^c$, $i_q^c$& 
Pre-disturbance active and reactive currents (pu) & 
1.0, 0 \\
$K_{pll}$& 
SRF PLL design: $k_p$ $k_i$ & 
0.025, 1.5 \\
$i_d^{c, ramp}$& 
Post fault active current ramp & 
28.4 kA/s \\
\hline
\end{tabular}
\label{tab1}
\end{table}

\begin{table}[h]
\caption{STATCOM PARAMETERS}
\label{table}
\setlength{\tabcolsep}{3pt}
\begin{tabular}{|p{40pt}|p{105pt}|p{70pt}|}
\hline
Symbol& 
Description& 
Value \\
\hline
$S_b $& 
Rated power& 
12 MVA \\
$V_g$& 
Nominal grid voltage (L-N, pk) & 
690 $\sqrt{2/3}$ V\\
$f_g$& 
Rated frequency & 
50 Hz \\
$r_{Lg}$, $L_{g}$& 
Transformer impedance & 0.25m$\Omega$, 0.018mH
\\
$i_d^c$, $i_q^c$& 
Active and reactive currents (pu) & 
0, 1.0 \\
$K_{pll}$& 
SRF PLL design: $k_p$ $k_i$ & 
0.025, 1.5 \\
\hline
\multicolumn{3}{p{240pt}}{Since the objective of the paper is to access stability and not the design, the STATCOM parameters are adapted from the WT system for simplicity.}
\end{tabular}
\label{tab1}
\end{table}

\subsection{Steady state load-flow}
In order to estimate the nonlinear stability boundary, i.e. the region of attraction, the equilibrium points of the system must be estimated. This can be simply obtained by solving the following equivalent nonlinear algebraic equations,  

\begin{equation}\label{}
\begin{aligned}
\begin{bmatrix}
v_{wt}\\
v_{st}
\end{bmatrix} &=
\begin{bmatrix}
Z_{11} & Z_{12}\\
Z_{21} & Z_{22}
\end{bmatrix}
\begin{bmatrix}
i_{wt}\\
i_{st}
\end{bmatrix} +
\begin{bmatrix}
v_{wt,0}\\
v_{st,0}
\end{bmatrix}
\end{aligned}
\end{equation}

The current injection from the aggregated WTs and the STATCOM is given by,
\begin{equation}\label{}
\begin{aligned}
i_{wt} &= (i_{d}^{c} cos x_1 - i_{q}^{c} sin x_1) \\
 &\qquad \qquad+j(i_{d}^{c} sin x_1 + i_{q}^{c} cos x_1) \\
i_{st} &= (- i_{st,q}^{c} sin y_1) +j(i_{st,q}^{c} cos y_1) 
\end{aligned}
\end{equation}
where,
\begin{equation}\label{}
\begin{aligned}
x_{1} &= sin^{-1} \big( \frac{R i_{q}^{wt} + w_n L i_{d}^{wt}}{|v_{wt}|}\big) + \angle v_{wt}\\
y_{1} &= sin^{-1} \big( \frac{R i_{d}^{st} }{|v_{st}|}\big) + \angle v_{st}
\end{aligned}
\end{equation}

The system equations (11)-(13) can be solved by the Newton method. The equilibrium points $x_{1,0}$ and $x_{3,0}$ will be later used to shift the initial RoA along the origin.

\subsection{Estimation of the RoA}
A hybrid approach to estimate the system RoA is employed \cite{13}. First, a linear energy function is established to estimate a closed set of initial conditions (i.e. initial RoA). Next, the reverse time trajectory technique is used to backward simulate all the initial conditions to obtain a closed time-limited region of attraction (TLRoA). 

\subsubsection{Initial conditions}
The first step in estimating the WT system boundary is obtaining an initial RoA, which is carried out by constructing a Lyapunov function (LF) from the linearised equations of the system (1) and (5). In \cite{17}, the system (1) was linearised around $\Tilde{\textbf{\textit{x}}}$, which gives
\begin{equation}
\begin{aligned}
\textbf{A}_{wt} &= \frac{\partial f_{wt}}{\partial x} |_{x = \Tilde{x}}\\
&=
\left[\begin{matrix}
  0 & 1\\
  \frac{\mp k_i V_g}{1- k_p L_g i_d^c} \sqrt{1-\gamma^2} & \frac{k_i L_g i_d^c \mp k_p V_g \sqrt{1-\gamma^2}}{1- k_p L_g i_d^c} 
\end{matrix}\right] 
\end{aligned}
\end{equation}

Similarly, the STATCOM (5) can be linearised as,
\begin{equation}
\begin{aligned}
\textbf{A}_{st} &= \frac{\partial f_{st}}{\partial x} |_{x = \Tilde{x}}\\
&=
\left[\begin{matrix}
  0 & 1\\
  \mp k_i V_s \sqrt{1-\gamma_s^2} & \mp k_p V_s \sqrt{1-\gamma_s^2} 
\end{matrix}\right] 
\end{aligned}
\end{equation}

If \textbf{A} = [$\textbf{A}_{wt}$, 0; 0, $\textbf{A}_{st}$] is Hurwitz matrix, then a quadratic LF $V(\textbf{\textit{x}})$ can easily be found as,
\begin{equation}\label{eq:8}
      V(\textbf{\textit{x}}) = \textbf{\textit{x}}^T\textbf{P}\textbf{\textit{x}}\\
\end{equation}
where, for any $\textbf{Q}\succ0$, $\textbf{P}\succ0$ is the solution of the Lyapunov equation,
\begin{equation}\label{eq:9}
      \textbf{P}\textbf{A} + \textbf{A}^T\textbf{P} + \textbf{Q} = 0\\
\end{equation}

By assuming, $\textbf{Q}$ as an identity matrix, $\textbf{P}$ can be computed from (17). Considering an energy level set of 0.001, the LF of the system obtained is,
\begin{equation}\label{eq:12z}
      V(\textbf{\textit{x}}) = ax_1x_2 + by_1y_2 + cx_1^2 + dx_2^2 + ey_1^2 + fy_2^2
\end{equation}
where, $a$=49.6, $b$=0.002, $c$=0.12, $d$=49.6, $e$=0.002, and $f$=0.12.

\begin{figure}[h]
    \centering
    \includegraphics[width=9.0cm]{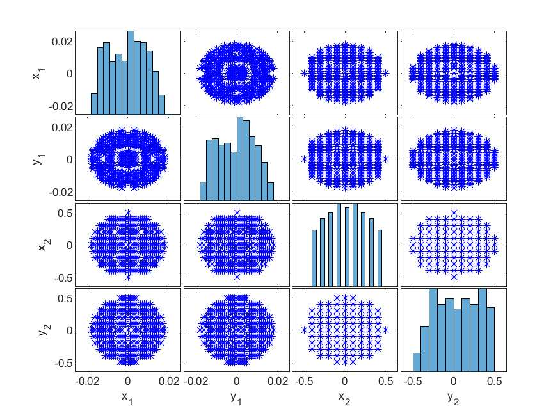}
    \caption{Projections of initial RoA estimate of the linearised model (1) and (5).}
    \label{fig:TrajectLRoa}
\end{figure}

Fig. \ref{fig:TrajectLRoa} illustrates the projections of the estimated 4D hyper ellipsoid (18) in 2D space. The hyper ellipsoid is the estimated initial RoA $A_{ini}(x)$, i.e., the boundary of this closed space gives us the set of initial conditions needed for the reverse time simulations. From \cite{13}\cite{15}, we know that if the RoA of the full system is $A(x_0)$, then,
\begin{equation}\label{}
    A_{ini}(x) \subset A(x_0)
\end{equation}
where, $x_0$ is the equilibrium point.

\subsubsection{Reverse time trajectory}
The time-reversal technique has been the subject of extensive research for several decades \cite{18}\cite{19}. If a system $\dot{x} = f(t, x)$ satisfies the Lipschitz condition, it guarantees a unique trajectory for each initial condition $x_i$. Similarly, solving the system backwards in time from $x_0$ results in traversing the same unique trajectory reaching $x_i$. This implies that the system $f$ can be solved as,

\begin{equation}\label{}
    \dot{x} = -f(t, x)\\
\end{equation}
with initial conditions chosen close to the equilibrium point.

A time-limited region of attraction, $A_{TL}(t, x)$, was introduced in \cite{13}, such that any trajectory on its boundary will reach the equilibrium at the same time, and any trajectory inside will reach the equilibrium in less than the said time. This implies that,
\begin{equation}\label{}
    A_{ini}(x) \subset A_{TL}(t, x) \subset A(x_0)
\end{equation}

For this study, the system (1)-(10) was solved backwards for 2.25 seconds, which is the considered time for the oscillations to dampen out. Figure \ref{fig:estimated TLRoA} illustrates the projections of the estimated 4D TLRoA in 2D space, where, the initial conditions can be obtained from $A_{ini}(x)$. It is important to note that when analysing the system boundary, the WTs and the STATCOM cannot be visualised in isolation, i.e. it is not enough to just analyse $x_1$ vs. $x_2$ and $y_1$ vs. $y_2$. Instead, all the states $x_1$, $x_2$, $y_1$ and $y_2$ must be analysed together.

\begin{figure}[h]
    \centering
    \includegraphics[width=9.0cm]{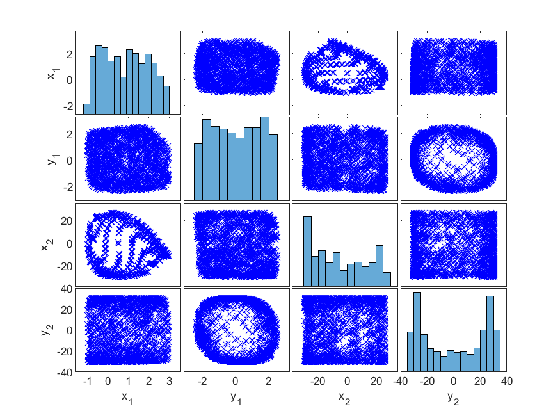}
    \caption{Estimated 4D TLRoA for the multi-converter system in the WPP: aggregated WTs and STATCOM.}
    \label{fig:estimated TLRoA}
\end{figure}

The TLRoA is a space inside which, if a trajectory enters, then it is guaranteed to attract towards the equilibrium point, however, this space can have islands or can be hollow, because of the homeomorphism property \cite{20}\cite{13}. This means that just the trajectory entering the space will not be a sufficient condition for stability, i.e., it is important to check for hollow space/islands. For example, Fig. 6 presents a 3D cup-like structure, i.e. a hollow cylinder with a base; its projections on a plane ($ax + by + cz = d$) are presented in Fig 6b. It can be observed that a 2D projection of the 4D space cannot reveal the hollow characteristics. A proper way of analysing such shapes would be to see cross-sections instead. Figure 6c presents the cross-section of the cup on the plane, which reveals that the shape is hollow.  

\begin{figure}[h]
    \centering
    \includegraphics[width=9.0cm]{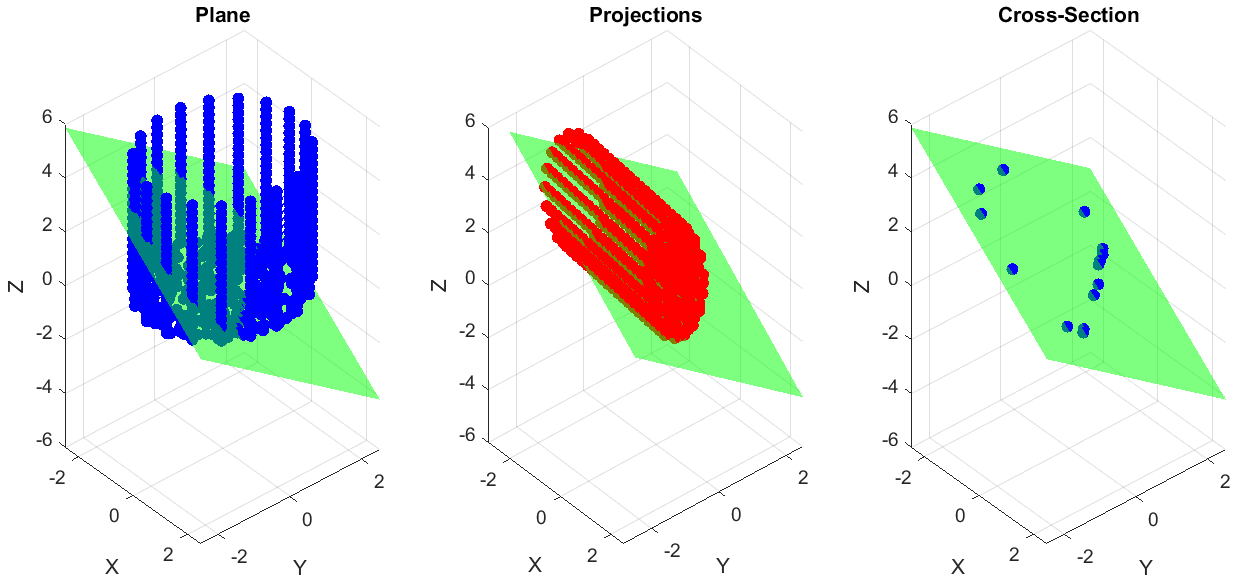}
    \caption{Visualisation of shape, (a) a 3D cup-like structure with a plane cutting it, (b) projection of the 3D cup on the plane, (c) cross-section of the 3D cup on the plane. A plane with a normal vector [1,1,1] and an intercept of 0.5 was selected.}
    \label{fig:estimated TLRoA}
\end{figure}

Similarly, the 4D TLRoA must be scoped for islands. Given the inherent challenge of visualising high-dimensional data, a proposed solution involves utilising a hyperplane (e.g., $ax + by + cz + dw = e$) to cut the estimated 4D TLRoA, and the 2D projections of the TLRoA's cross-section can be visualised. Various hyperplanes are tested, but for an illustrative purpose, the 2D projections of the TLRoA's cross-section on a hyperplane with a normal vector [1,1,1,1] and an intercept of 30 are presented in Fig. 7. It must be noted that the number of sample points in the cross-section on the hyperplane influences the clear visibility of islands. More sample points give a better representation, but a trade-off must be carried out to select the number of sample points versus the computation burden to estimate the TLRoA. This will be covered as a future scope. In this work, only the outside boundary of the estimated TLRoA is validated.    

\begin{figure}[h]
    \centering
    \includegraphics[width=8.0cm]{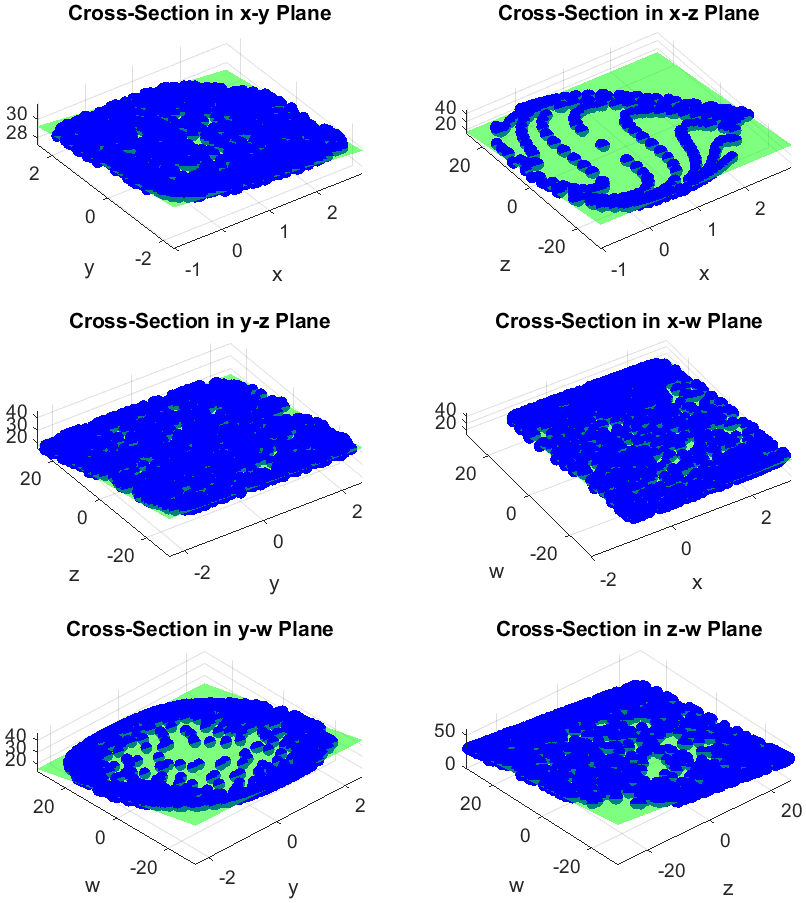}
    \caption{2D projections of the TLRoA's cross-section when cut with a hyperplane with a normal vector [1,1,1,1] and an intercept of 30.}
    \label{fig:estimated TLRoA1}
\end{figure}

\section{Time domain verification}
In this section, we assessed our methodology for nonlinear stability boundary through time-domain simulations employing an EMT WPP switching model in PSCAD. The EMT model used the configuration specified in \cite{16}, wherein the current controller gains were tuned to achieve a fast response. 

\subsection{Estimation of Critical clearing time}
In the context of RoA, the Critical Clearing Time (CCT) represents the duration required for the system to traverse from its equilibrium point to the RoA boundary. It serves as a crucial metric, defining the maximum allowable time within which a disturbance must be mitigated to ensure the system's stability.
The CCT is usually obtained for a specific disturbance. For instance, in this work, for illustrative purposes, the clearing time for a bolted three-phase to-ground fault at the connection point of the WPP is investigated. A similar approach can be carried out to investigate other faults and fault locations.  

\begin{figure}[h]
    \centering
    \includegraphics[width=8.50cm]{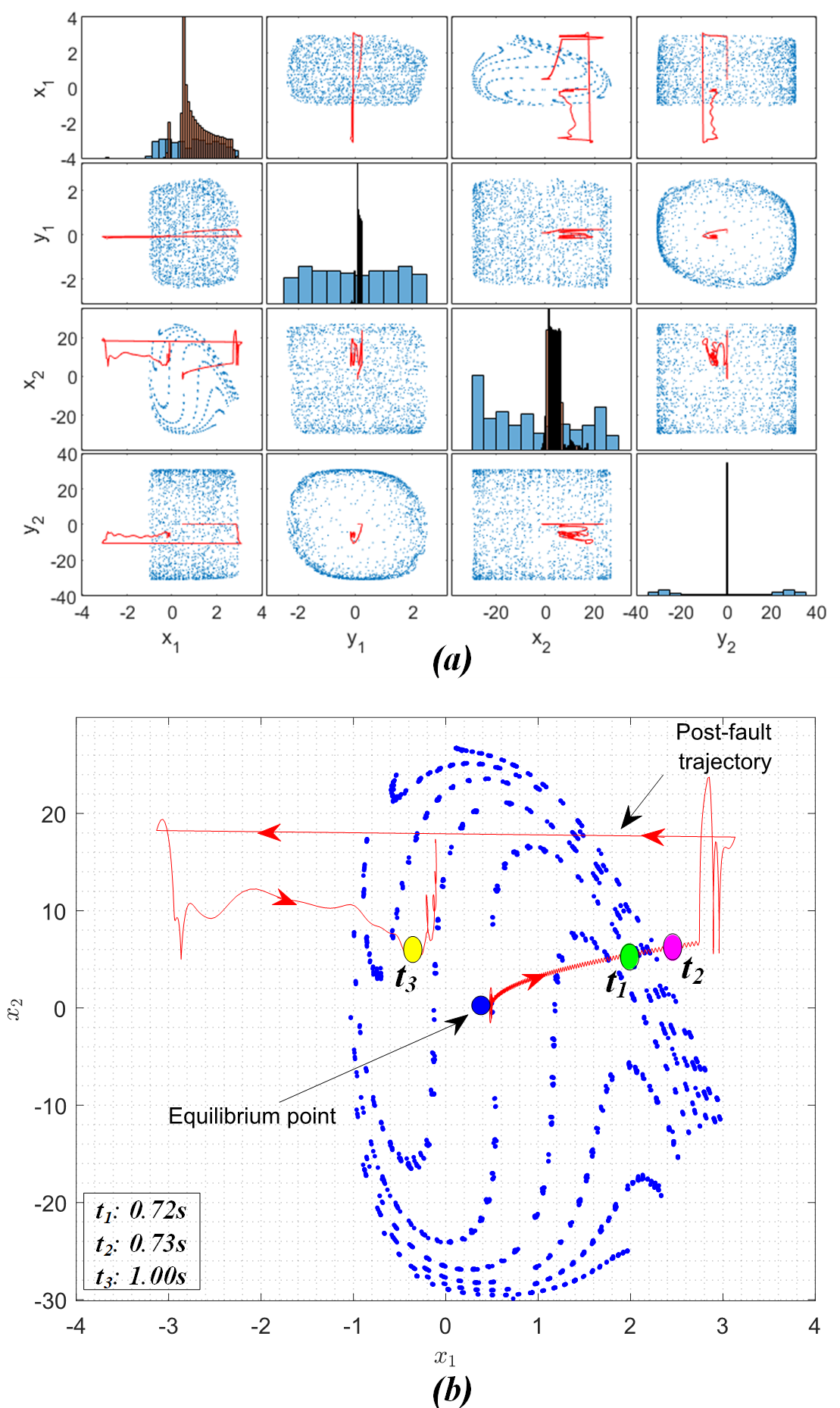}
    \caption{Estimation of clearing time: (a) Fault trajectory against estimated 4D TLRoA, (b) Zoomed in $x_1$ vs $x_2$ plot.}
    \label{fig:estimated_TLRoA2}
\end{figure}

The methodology outlined in \cite{13} and \cite{15} is applied for the estimation of clearing time. In the PSCAD environment, a sustained bolted three-phase to-ground fault is simulated at the WPP's connection point for an extended duration, specifically 1.2 seconds. Subsequently, the fault trajectory is plotted against the estimated TLRoA, as depicted in Figure \ref{fig:estimated_TLRoA2}(a). Notably, in Figure \ref{fig:estimated_TLRoA2}(a), it is evident that the fault trajectory breaches the TLRoA boundary only the concerning states of WTs (i.e. $x_1$ and $x_2$). Consequently, a zoomed view of the $x_1$ vs $x_2$ plot is presented in Figure \ref{fig:estimated_TLRoA2}(b). In this zoomed-in representation, it is clear that the fault trajectory intersects the TLRoA boundary multiple times, signifying instances of exit and re-entry into the TLRoA. A more detailed exploration of this phenomenon is presented in \cite{15}, where it was discussed that as long as the fault trajectory is cleared inside the TLRoA, the system will be stable. In this regard, the fault is cleared at times $t_1$, $t_2$ and $t_3$ (0.72 s, 0.73 s and 1 s), and the stability of the system is investigated. 

\begin{figure}[h]
    \centering
    \includegraphics[width=9.0cm]{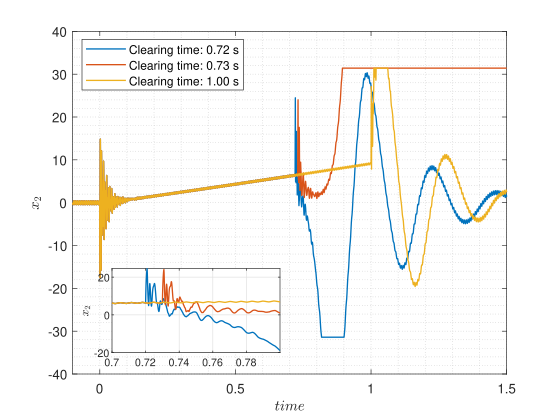}
    \caption{Verification of the identified critical clearing time by PSCAD simulations.}
    \label{fig:estimated_TLRoA3}
\end{figure}

To validate the accuracy of our estimated clearing times, from PSCAD, the time-domain trajectory of the WTs state $x_2$ is presented in Fig. \ref{fig:estimated_TLRoA3}. The system stability at the clearing times is in line with the theoretical understanding; this validation process reinforces the credibility and reliability of our proposed nonlinear stability assessment methodology.

\subsection{Discussion}
This research focuses on the estimation of nonlinear stability boundaries for aggregated WPPs, emphasising on low-frequency converter dynamics. Nevertheless, it's crucial to acknowledge that aggregating WTs is not universally applicable, especially in cases where the network topology is asymmetric, involving variations in power output or cable lengths. In such scenarios, aggregating these WTs may necessitate the representation of the aggregated unit by two or more machine equivalents.
One notable advantage of our proposed methodology is its ability to obtain the nonlinear boundaries for such complex topologies using the reverse time trajectory approach. However, the task becomes progressively challenging when visualising the TLRoA in a lower-dimensional space. Simultaneously, the number of initial points for reverse time simulations increases significantly, necessitating the implementation of adaptive sampling techniques; this will be covered as a future scope of work. It must be noted that these techniques must maintain a smooth boundary representation, enabling the visualisation of potential islands within the RoA, and managing the computational burden efficiently.

\section{Conclusion}
This work presents a novel methodology for WPP stability analysis. Our primary contribution lies in the derivation of nonlinear stability conditions, specifically defining the stable region for a WPP featuring aggregated WTs, a STATCOM, and an extensive collector and transmission cable network.
The fundamental contribution of our work is presenting an n-dimensional stability boundary/ time-limited region of attraction (TLRoA). This boundary, which exists within the broader RoA of the system, has been obtained through the application of reverse time simulations. Notably, the TLRoA allows us to better understand the dynamic behaviour of complex power systems when subjected to disturbances.

One of the merits of our methodology is its ability to move beyond the constraints of conventional analytical approaches. These traditional methods often struggle to scale effectively for large wind power plants. In contrast, our approach demonstrates remarkable versatility and applicability to the analysis of system stability under large disturbances, regardless of the size or complexity of the WPP.
To underscore the credibility and reliability of our findings, we compare our results with an established benchmark model. This alignment enhances the robustness of our approach and, most importantly, reinforces the assurance of power system stability.


%





\ifCLASSOPTIONcaptionsoff
  \newpage
\fi

\begin{IEEEbiography}[{\includegraphics[width=1in,height=1.25in,clip,keepaspectratio]{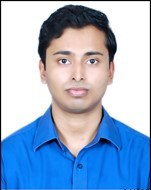}}]{Sujay Ghosh} Sujay Ghosh is currently pursuing his Ph.D. on stability of large renewable power plants in low inertia systems, from Technical University of Denmark (DTU) in collaboration with Ørsted Wind power A/S. He has a Masters's in Science in Electrical engineering from DTU. His research field is in power system stability and control. Previously, he has also worked as a system studies engineer in the electrical industry in India and Denmark.
\end{IEEEbiography}

\begin{IEEEbiography}[{\includegraphics[width=1in,height=1.25in,clip,keepaspectratio]{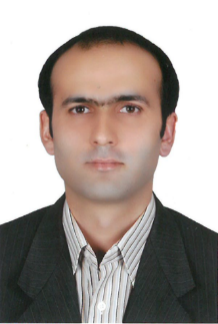}}]{Mohammad Kazem Bakhshizadeh} received the B.S. (2008) and M.S. (2011) degrees in electrical engineering from Amirkabir University of Technology, Tehran, Iran, and the PhD degree from Aalborg University, Aalborg, Denmark in 2018. In 2016, he was a visiting scholar at Imperial College London, London, U.K.
He is currently a senior power system engineer at Ørsted Wind Power, Fredericia, Denmark. His research interests include power quality, modelling, control and stability analysis of power converters, and grid converters for renewable energy systems.
\end{IEEEbiography}

\begin{IEEEbiography}[{\includegraphics[width=1in,height=1.25in,clip,keepaspectratio]{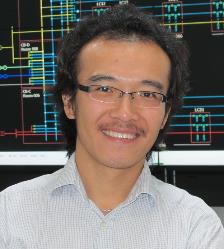}}]{Guangya Yang} (M'06--SM'14) Guangya Yang joined Technical University of Denmark (DTU) in 2009 and currently senior scientist with the Department of Wind and Energy Systems. Previously he has also been a full-time employee at Ørsted, working on electrical design of large offshore wind farms. His research field is in stability, control, and operation of power systems. He has received numerous research grants, including the recent H2020 MSCA ITN project InnoCyPES as coordinator. He is currently serving as lead editor for the IEEE Access Power and Energy Society Section and has been an editorial board member of several journals. He is a member of the IEC Technical Committee “Wind Power Generation Systems” (TC88) and the convenor of IEC61400-21-5. 
\end{IEEEbiography}

\begin{IEEEbiography}[{\includegraphics[width=1in,height=1.25in,clip,keepaspectratio]{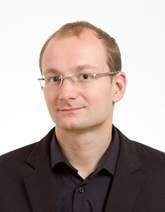}}]{Lukasz Kocewiak} (M’12–SM’16) Łukasz Hubert Kocewiak received the BSc and MSc degrees in electrical engineering from Warsaw University of Technology in 2007 as well as the PhD degree from Aalborg University in 2012. Currently, he is with Ørsted and is working as an R\&D manager. He is a power system specialist in the area of design of electrical infrastructure in large offshore wind power plants. The main direction of his research is related to harmonics, stability and nonlinear dynamics in power electronics and power systems especially focused on wind power generation units. He is the author/co-author of more than 100 publications. He is a member of various working groups/activities within Cigré, IEEE, and IEC.
\end{IEEEbiography}




\end{document}